\documentclass[letterpaper, 10 pt, conference]{ieeeconf}  

\IEEEoverridecommandlockouts                              

\usepackage{microtype}                 
\PassOptionsToPackage{warn}{textcomp}  
\usepackage{textcomp}                  
\usepackage{mathptmx}                  
\usepackage{times}                     
\usepackage{cite}                      
\usepackage{tabu}                      
\usepackage{booktabs}                  
\usepackage{graphicx} 
\usepackage{caption}
\usepackage{subcaption}
\usepackage{float}
\usepackage{cleveref}
\let\labelindent\relax
\usepackage{enumitem}
\usepackage{url}
\setlist{nolistsep}
\title{\LARGE \bf
Effects of Augmented-Reality-Based Assisting Interfaces on Drivers' Object-wise Situational Awareness in Highly Autonomous Vehicles
}

\author{Xiaofeng Gao$^{*}$\thanks{$^{*}$Xiaofeng Gao is with the University of California, Los Angeles. e-mail: xfgao@ucla.edu. The work was done at Honda Research Institute USA.}
\and Xingwei Wu$^{\dagger\ddagger}$\thanks{$^{\dagger}$Xingwei Wu, Samson Ho, Teruhisa Misu and Kumar Akash are with the Honda Research Institute USA. e-mail:\{sho, tmisu, kakash\}@honda-ri.com} \thanks{$^{\ddagger}$Xingwei Wu is currently with Zoox. e-mail: xingweiwu19@gmail.com. }
\and Samson Ho$^{\dagger}$ \and Teruhisa Misu$^{\dagger}$ \and Kumar Akash$^{\dagger}$
}

\begin{document}

\maketitle
\thispagestyle{empty}
\pagestyle{empty}

\begin{abstract}

Although partially autonomous driving (AD) systems are already available in production vehicles, drivers are still required to maintain a sufficient level of situational awareness (SA) during driving. Previous studies have shown that providing information about the AD's capability using user interfaces can improve the driver's SA. However, displaying too much information increases the driver's workload and can distract or overwhelm the driver. Therefore, to design an efficient user interface (UI), it is necessary to understand its effect under different circumstances. In this paper, we focus on a UI based on augmented reality (AR), which can highlight potential hazards on the road. To understand the effect of highlighting on drivers' SA for objects with different types and locations under various traffic densities, we conducted an in-person experiment with 20 participants on a driving simulator. Our study results show that the effects of highlighting on drivers' SA varied by traffic densities, object locations and object types. We believe our study can provide guidance in selecting which object to highlight for the AR-based driver-assistance interface to optimize SA for drivers driving and monitoring partially autonomous vehicles.

\end{abstract}


\section{Introduction}
Autonomous vehicles (AVs) have the potential to revolutionize the transportation industry. Despite the rapid development of the autonomous driving (AD) system, fully automated cars are still not available on public roads. Currently, some vehicles on the market are equipped with advanced driver assistance systems (ADAS) that allow partially automated driving, or SAE Level 2 (L2) automation \cite{sae2018taxonomy}. While drivers can briefly enjoy feet-free and hands-free driving under certain driving situations at this level of automation, they are still required to monitor the traffic conditions and prepare for sudden maneuvers and possible takeover requests. As a result, it is crucial to maintain the driver's situational awareness (SA) when interacting with the AD system and avoid the out-of-the-loop problem \cite{endsley1988design}. 

With the goal of improving drivers' SA and trust, researchers investigated various ways of communication to convey internal information to the drivers. The challenge is that showing additional information to the drivers can increase their cognitive load and cause distractions \cite{helldin2014transparency}. Showing too much information not only prevents drivers from paying attention to the most critical information during driving \cite{ananny2018seeing}, but is also against the main motivation of developing the AD system, i.e. reducing driver workload. Therefore, we believe that a smart user interface (UI) should be able to strike a balance between the amount of information provided and the driver's limited attention and cognitive load.

\begin{figure}[t]
    \centering
    \includegraphics[width=0.49\textwidth]{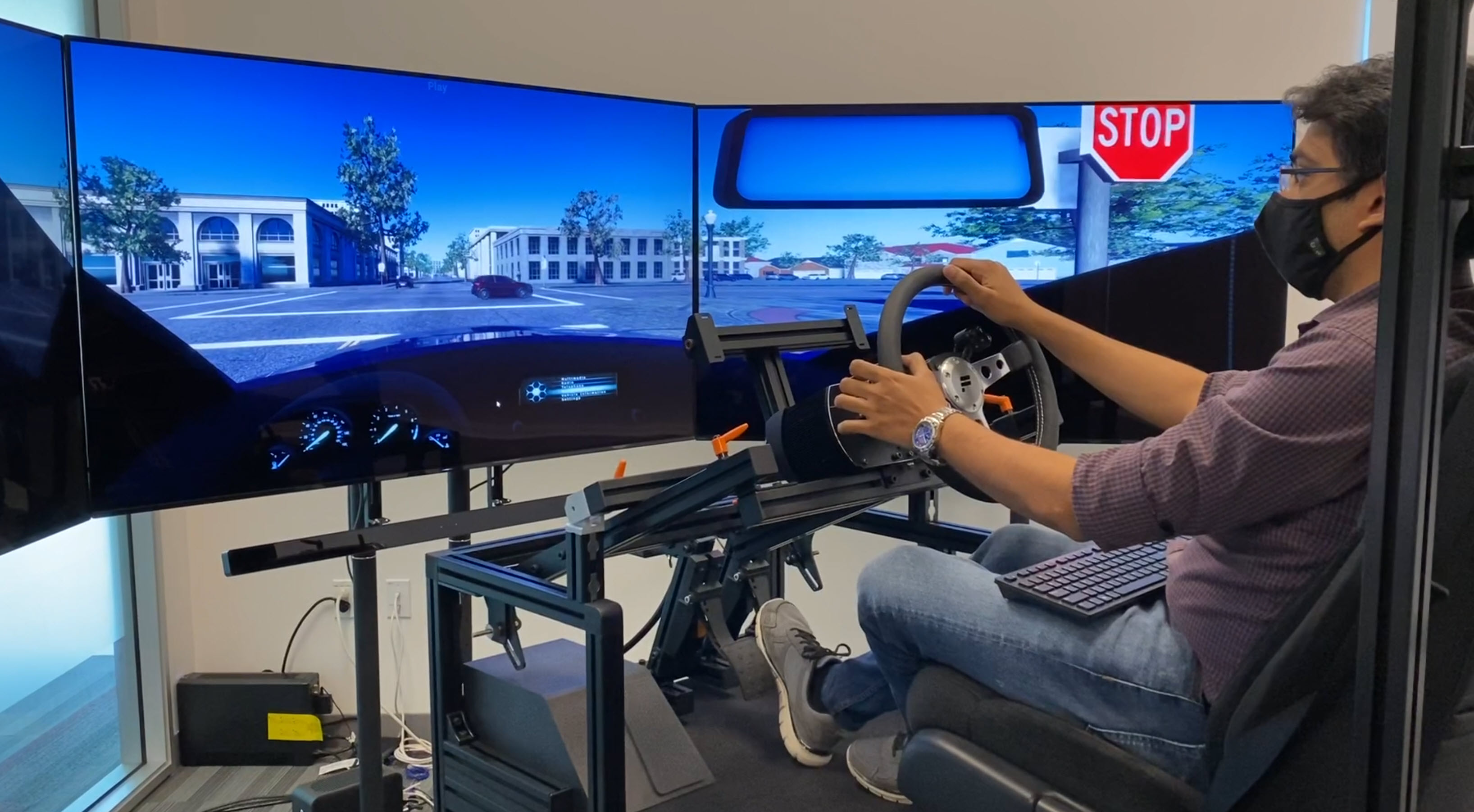}
    \caption{Our driving simulator is composed of a steering wheel and two pedals mounted on a cockpit, and three 55-inch displays showing the front and side views of the virtual environment.}
    \label{fig:simulator}
\end{figure}

Results from previous studies showed that highlighting hazardous objects via augmented reality (AR) based UI is a promising way to increase drivers' SA \cite{lindemann2018catch, colley2021effects}. Nevertheless, those works mainly focused on evaluating the effects of highlighting on SA \textbf{across all objects}. We believe a smart UI should optimize the highlighting for each object to maintain a proper workload and help the driver be aware of the potential hazards that are prone to be ignored. Therefore, we need to understand the effects of highlighting on \textbf{each specific object}, depending on the object characteristics. 

Specifically, in this work, we distinguish objects by three properties 1) locations (relative to the driver), 2) types (i.e. pedestrian or vehicle), and 3) traffic densities and evaluate the effect of highlighting considering those factors.

We implemented object highlighting via a UI on a driving simulator based on Unreal Engine 4 (UE4), and conducted an in-person study (N=20) on the simulator to investigate the effect of highlighting on drivers' attention allocation and SA for each object in an urban environment.   

\begin{figure*}[t]
\centering
\includegraphics[width=\textwidth]{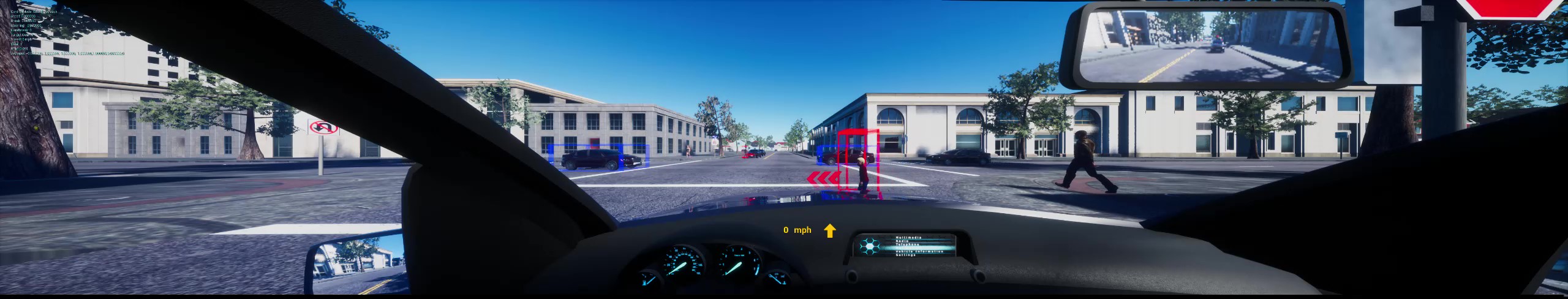}
\caption{This is a forward event intersection with high traffic density, corresponding to the event in \Cref{fig:events_f}. We highlight objects using bounding boxes on the user interface: red for pedestrians and blue for cars. In addition, we also display the ego vehicle's current speed and heading direction with yellow texts and arrows in the middle. During the study, this concatenated screenshot is separately shown on three displays to simulate the field-of-view of a driver in the real world (see \Cref{fig:simulator}).}
\label{fig:user_interface}
\end{figure*}

\begin{figure*}[t]
\centering
\includegraphics[width=\textwidth]{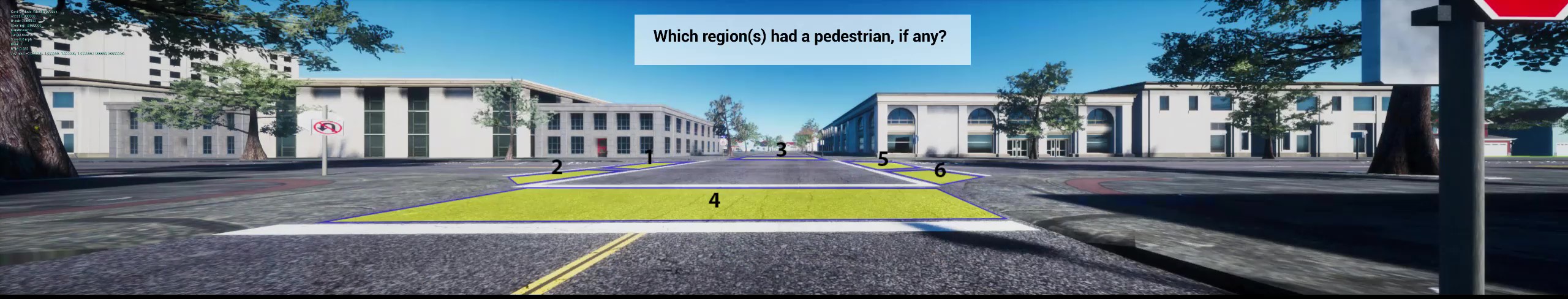}
\caption{To evaluate drivers' SA, we pause the simulation and hide all road users. On top of the background road scene of the intersection, we display several regions and ask users to choose which regions were occupied by pedestrians or vehicles.}
\label{fig:sagat}
\vspace{-10pt}
\end{figure*}

The main contributions of this paper are:
\begin{itemize}[leftmargin=*]
    \item We implemented an AR-based UI in a driving simulator to inform drivers of the AD's perception capabilities by highlighting hazardous objects. We focused on urban intersections because of their complex traffic conditions, which can be demanding for the drivers to monitor.  
    \item We designed and conducted a simulator experiment to evaluate the impact of highlighting on drivers' object-wise SA and attention allocation. Specifically, we designed a novel Situational Awareness Global Assessment Technique (SAGAT) protocol with temporal variations to measure the same driver's SA changes before and after the highlighting in two identical intersections to better understand the effect of highlighting on SA. Objects' locations and movements at intersections are discretized based on spatial distance and eccentricity.  
    \item We carefully analyzed the effect of highlighting with the AR interface compared to non-highlighting in different conditions, including a combination of object types, object locations and traffic densities. 
\end{itemize}
The results of our study suggest that the effects of highlighting on perception-level SA highly depend on object properties and traffic densities. We believe the results pave the way for a smart UI that can selectively highlight objects to improve SA for drivers of AVs, leading to more safety in driving and monitoring partially autonomous vehicles.

\section{Related Work}
\subsection{Situational Awareness Measurements}
SA can be generally understood as knowing what is going on around you \cite{endsley1988design}. Over the years, various methods have been proposed to measure SA. They can be categorized into objective measurements (e.g. SPAM \cite{durso1998situation} and DAZE \cite{sirkin2017toward}) and subjective measurements (e.g. SART \cite{taylor1990situational} and SARS \cite{waag1994tools}). SAGAT is a widely-known technique to measure SA objectively \cite{endsley1988situation}. During a SAGAT session, the display is frozen at selected times and participants are asked to answer questions to measure SA. An advantage of SAGAT is that participants are unable to prepare for the questions in advance, thus minimizing the possibility of attention bias. Studies suggested that SAGAT is a technique with a high degree of reliability \cite{endsley1994individual} and validity \cite{endsley1990predictive}. Apart from direct measurements, SA can also be inferred with indirect measurements, including eye gaze behaviors and takeovers \cite{barnard2010spotting, yang2018hmi,Zhu2021ImprovingDS}. For a more comprehensive review of situational awareness measurements, we refer readers to this survey \cite{salmon2006situation}.

\begin{figure*}[t]
\centering
\begin{subfigure}{0.64\columnwidth}
    \centering
    \includegraphics[width=\columnwidth]{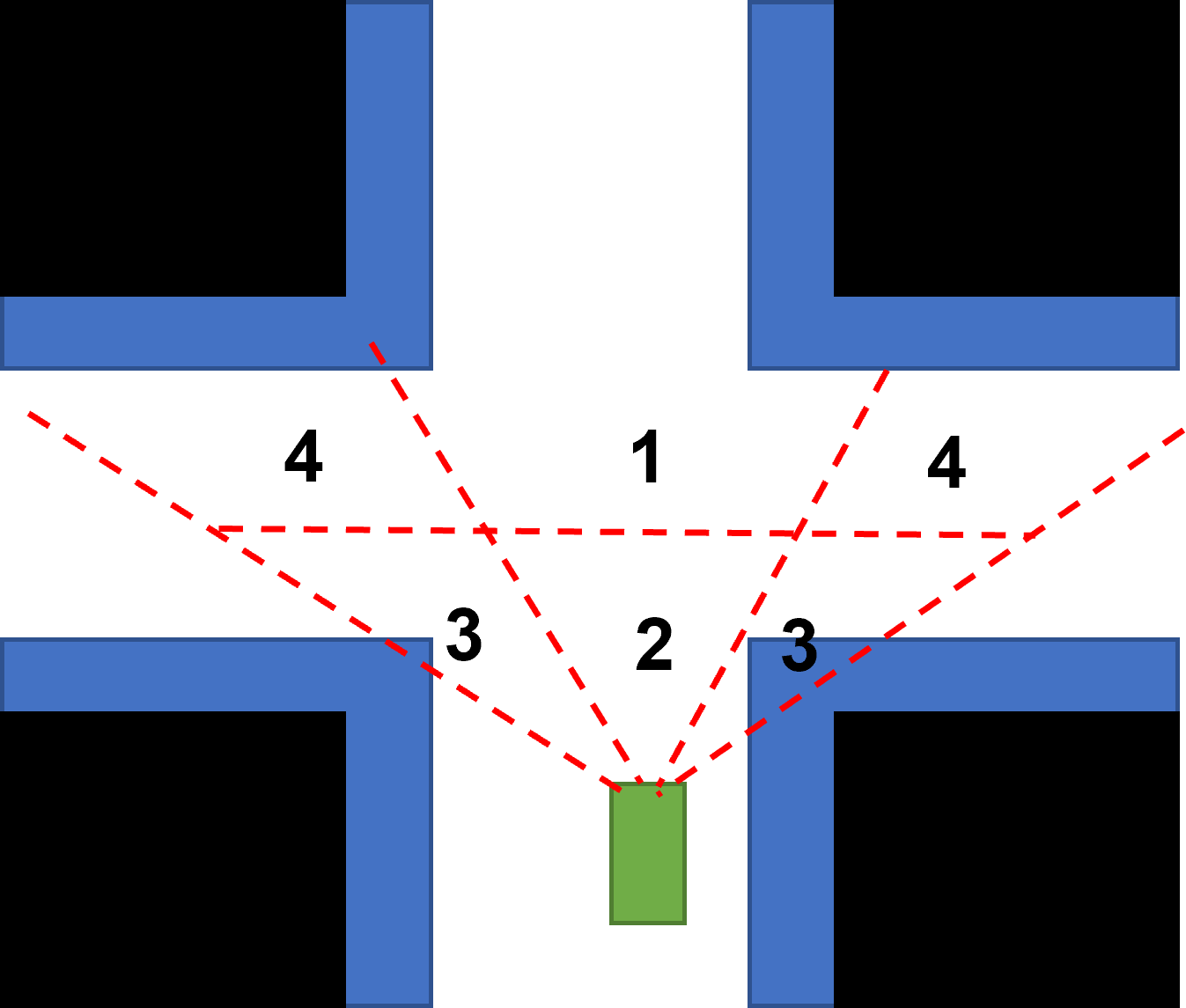}
    \caption{Area Discretizations based on spatial distance and eccentricity relative to the ego vehicle (green).}
    \label{fig:area_dis}
\end{subfigure}
\hfill
\begin{subfigure}{0.64\columnwidth}
    \centering
    \includegraphics[width=\columnwidth]{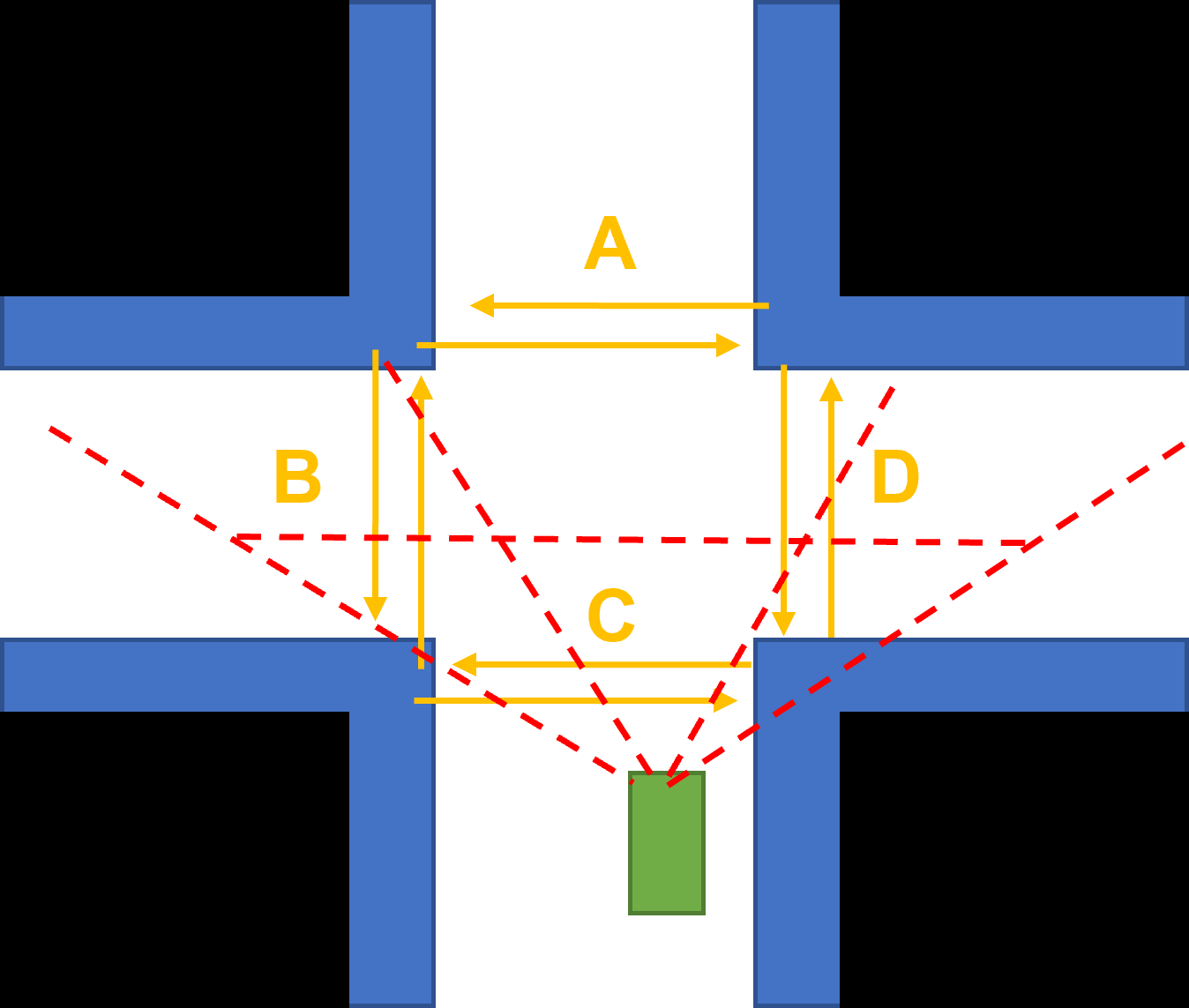}
    \caption{Pedestrian movements. Yellows arrows show target pedestrians' all possible movements across the intersection.}
    \label{fig:ped_dis}
\end{subfigure}
\hfill
\begin{subfigure}{0.64\columnwidth}
    \centering
    \includegraphics[width=\columnwidth]{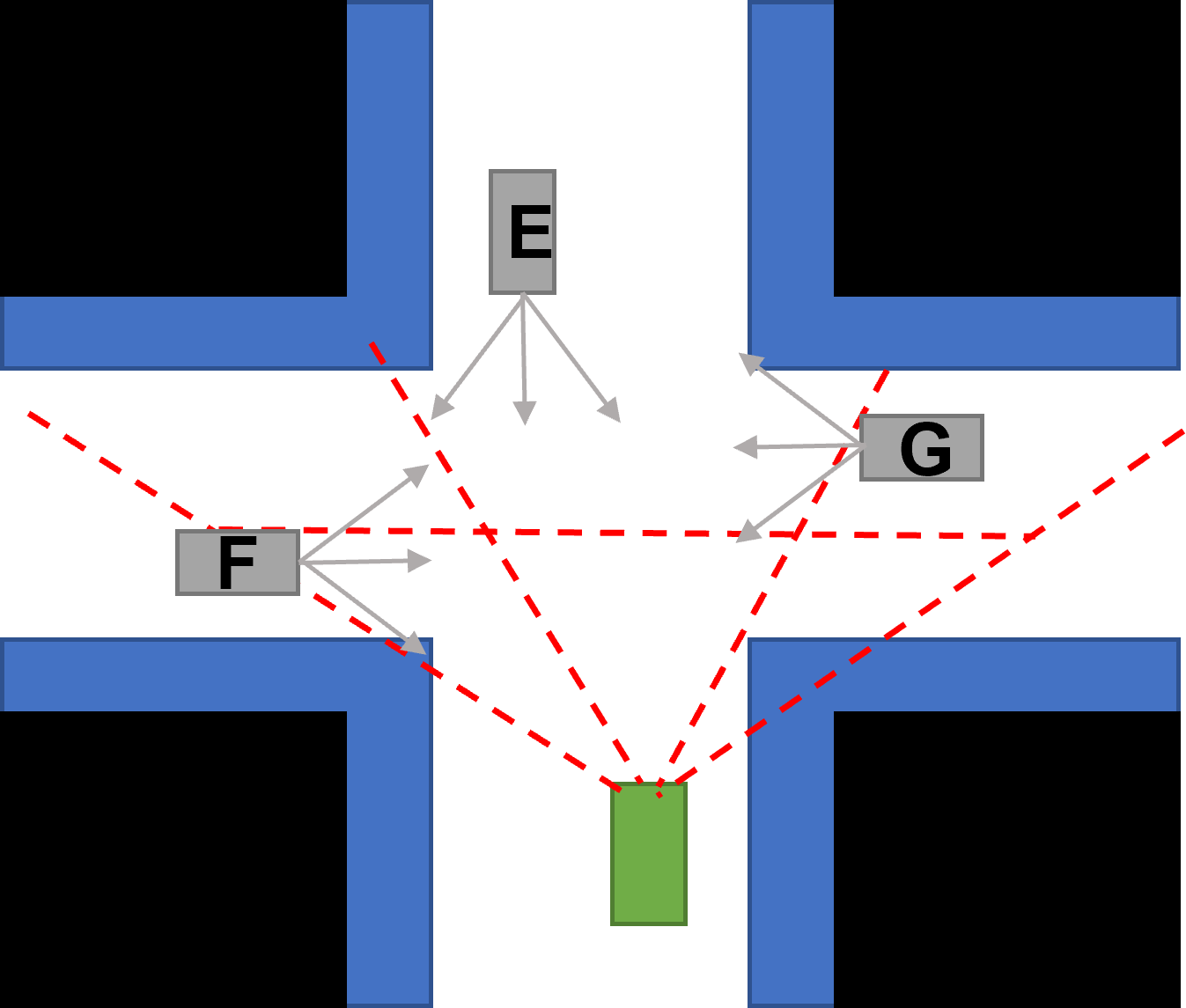}
    \caption{Vehicles movements. Gray rectangles are target vehicles' initial locations and arrows show vechicles' possible moving directions.}
    \label{fig:veh_dis}
\end{subfigure}

\caption{We first discretize an intersection into 4 areas based on spatial distance and eccentricity relative to the green ego vehicle. According to area discretizations, we then discretize pedestrians' and vehicles' all possible movements near the intersection. For example, pedestrian A crossing the top of the intersection is considered moving in area 1, while car F going straight on the left will be moving in areas 2 and 3.}
\label{fig:obj_discretization}
\vspace{-10pt}
\end{figure*}

\subsection{Ways to Improve Driving Situational Awareness}
Driver’s SA at SAE L2 or L3 automated driving has been widely studied for years. Previous works have shown that driver’s SA can be influenced by a wide range of factors, such as age \cite{bolstad2001situation}, driving experience \cite{wright2016experienced} and working memory \cite{johannsdottir2010role, heenan2014effects}. Since driver’s SA plays an important role in driving safety, different methods have been proposed to enhance SA during driving. Recently, the idea of assisting human-machine interface (HMI) has generated much interest \cite{merenda2018augmented}. Studies that examined the effects of AR windshield display (WSD) interface found significant effects on driver's SA, when highlighting potential driving threats \cite{lindemann2018catch} or common traffic objects (e.g. cars, pedestrians, and traffic signs) \cite{colley2021effects,wang2020situation}. The SA improvement was observed at both perception level \cite{tong2019augmented} and comprehension and projection levels \cite{phan2016enhancing}. In these works, the focus is to evaluate the effect of HMI on the driver's average SA across all traffic objects between experimental groups. Thus the driver's SA is measured within each treatment group without distinguishing between objects \cite{phan2016enhancing, lindemann2018catch, colley2021effects}. Our work takes one further step in this direction: 1) we focus on evaluating the effect of HMI on driver's SA for each object, distinguished by their locations and types, and 2) propose a novel SAGAT protocol with temporal variations by measuring the same participant's SA before and after the treatment to better analyze its effects toward a specific object.


\section{Method}
In this section, we introduce our simulation study. We start with participant and apparatus in \Cref{sec:participants} and talk about the AD system and AR cues in \Cref{sec:AD}. We discuss how we discretize the object locations in \Cref{sec:discretization} and the details of the scenarios in \Cref{sec:scenarios}. Then we describe how we measure attention allocation and situational awareness of the driver in \Cref{sec:dependent}. Finally, we go through the whole procedure of our study in \Cref{sec:procedure}.

\subsection{Participants and Apparatus}
\label{sec:participants}
A total of 20 participants (12 males, and 8 females) from the San Francisco Bay area completed the study. Their age ranged from 20 to 49 years old. To be eligible for the study, each participant was required to have had a valid license for more than two years and drive more than 5,000 miles (8,047 km) per year. 

As shown in \Cref{fig:simulator}, we used a medium-fidelity driving simulator built with AirSim \cite{shah2018airsim}, a plug-in for UE4, to conduct all driving sessions. Also, Tobii Pro Glasses 3\footnote{\url{https://www.tobiipro.com/glasses3}} were used to collect the participant's eye-tracking data. 

\subsection{SAE L2 AD System and AR Assisting Cues}
\label{sec:AD}
The Wizard of Oz method is used to simulate realistic AD driving. To more realistically emulate a functioning AD system, we had an expert driver drive the ego-vehicle through the premeditated route in the simulated environment. The drive was recorded by saving all pedal and steering inputs to an AD file, which then provided realistic autonomous car behavior to participants. To ensure consistency of the AD driving behavior, all driving data were recorded from a single expert driver. The driving behaviors were further reviewed by two researchers, and routes were practiced to ensure consistency. During the driving, the participant was requested to indicate their take-over intention by pressing the brake pedal. 

The AR cues were developed and assigned to highlight specific road users within intersections. As shown in \Cref{fig:user_interface}, AR graphics used for highlighting were consistent in shape, i.e. a series of 3D bounding boxes that formed a cubic region surrounding a specific type of road users ({\it blue} for vehicles and {\it red} for pedestrians). 

\begin{figure*}[t]
\centering
\includegraphics[width=0.9\textwidth]{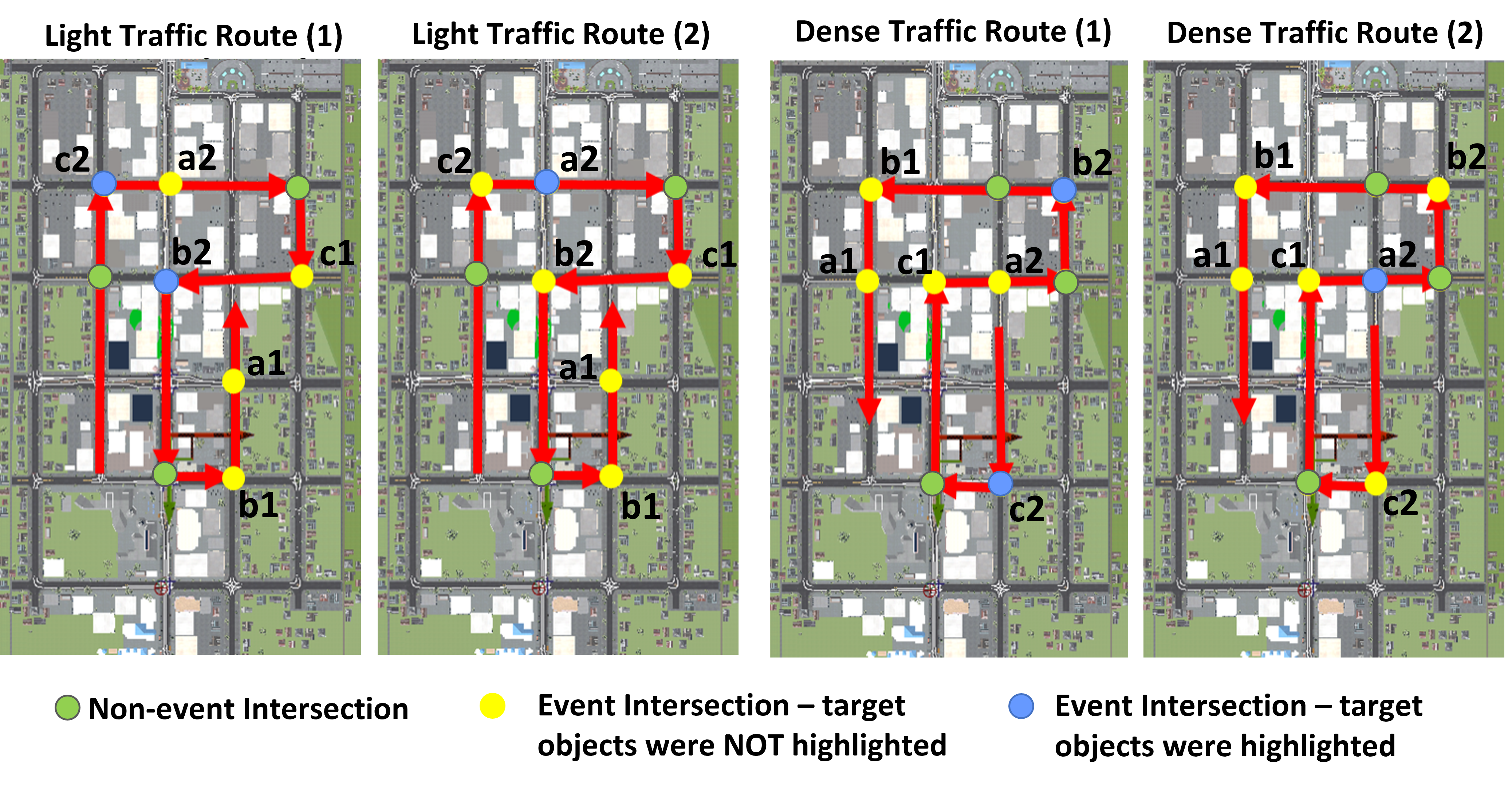}
\caption{Drives and intersections. Light traffic route 1 (LT1) and light traffic route 2 (LT2) are of low traffic density, while dense traffic route 1 (DT1) and dense traffic route 2 (DT2) are of high traffic density. Blue dots represent event intersections where the target objects are highlighted. Yellow dots represent event intersections where the target objects are not highlighted. Green dots are non-event intersections where we ask dummy SAGAT questions to reduce the learning effect of SA in event intersections. In intersections a1, b1 and c1, SAGAT questions are asked before the treatment (highlighting or not highlighting). In intersections a2, b2 and c2, SAGAT questions are asked after the treatment.}
\label{fig:drives}
\vspace{-10pt}
\end{figure*}

\subsection{Object Location Discretization}
\label{sec:discretization}
The human visual field can be commonly divided into three major regions: foveal, parafoveal, and peripheral. The foveal region extends out to an angle of 1 degree and the parafoveal region from 1 to 5 degrees \cite{nelson1980functional, quinn2019clinical}. Those two together are commonly referred to as the central vision, and the peripheral region encompasses the remainder of the visual field. Many researchers have noted that, as a result of the inhomogeneity of the visual system, attention allocation and awareness are strongly affected by the target eccentricity and spatial distance. Detecting a target far away in the peripheral as opposed to nearby and central vision requires longer search times and more eye movements \cite{carrasco1995eccentricity}.  

Considering both an object's spatial distance (i.e. near or far) and eccentricity (i.e. center or marginal) relative to the driver in the ego vehicle, we categorized the object positions in an intersection into four types of areas: the top center area (area 1), the bottom center area (area 2), the bottom left and bottom right areas (area 3) and the top left and top right areas (area 4), as shown in \Cref{fig:area_dis}. Since the pedestrians and cars move across different areas, their movements were also discretized (\Cref{fig:ped_dis}, \Cref{fig:veh_dis}).

\subsection{Driving Scenario and Events} 
\label{sec:scenarios}
The driving scenario is an urban environment in daylight conditions with a posted speed limit of 25 mph. Events are triggered when the ego vehicle comes near event intersections. During an event, the ego vehicle first stops before the intersection due to a stop sign or a flashing red traffic light. The vehicle then waits until the other road users have passed the intersection following the traffic rule. While the vehicle is waiting, the driver is asked to continuously monitor the surroundings and take over the control if the AD system has made an unexpected or dangerous move. The cars and pedestrians across intersections are consistent in appearance. Intersections are of similar sizes ($L=15.3 \pm 2.0$ m, $W=14.4 \pm 1.2$ m).

\begin{figure*}[t]
\centering
\begin{subfigure}{0.64\columnwidth}
    \centering
    \includegraphics[width=\columnwidth]{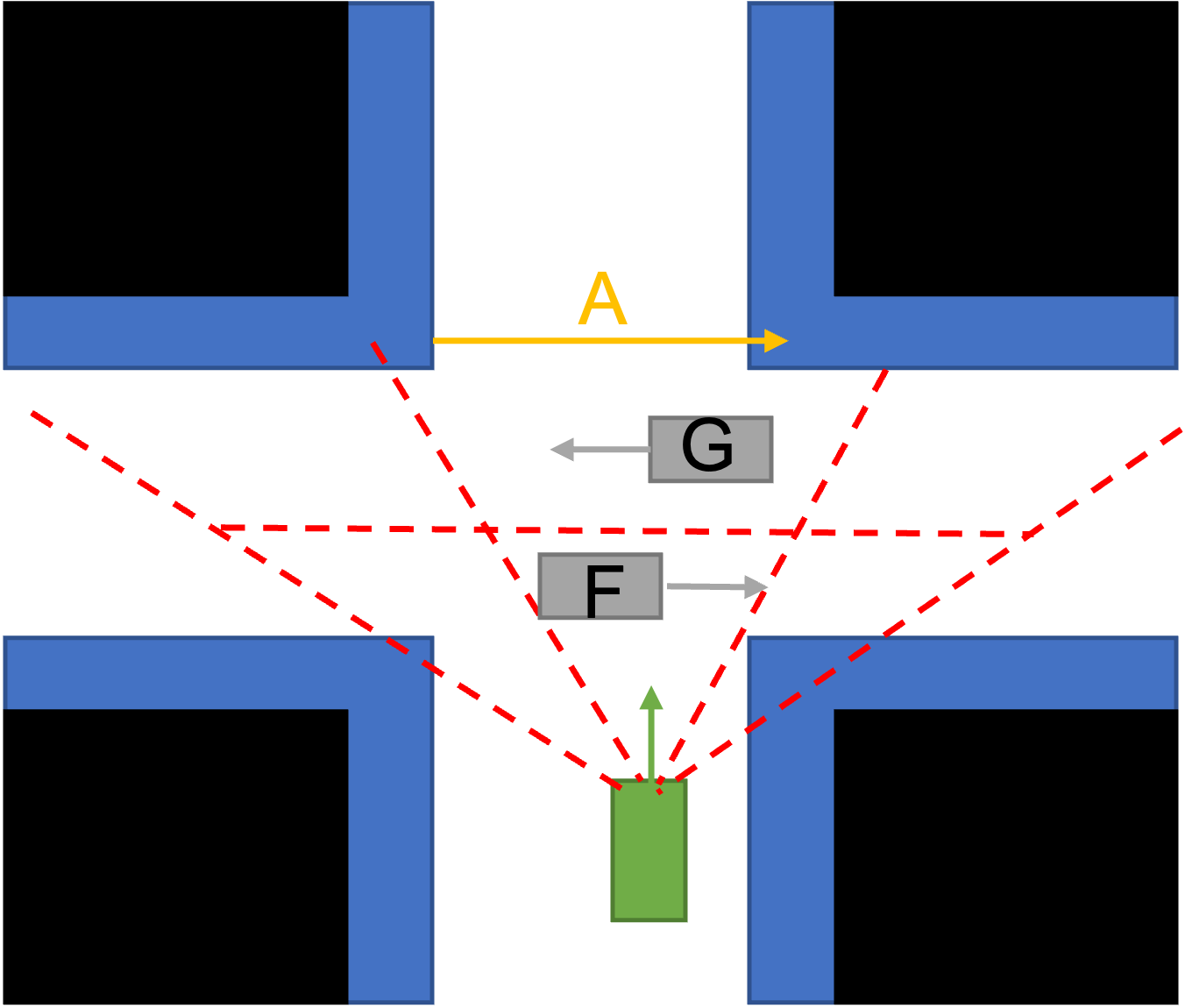}
    \caption{Forward event intersection}
    \label{fig:events_f}
\end{subfigure}
\hfill
\begin{subfigure}{0.64\columnwidth}
    \centering
    \includegraphics[width=\columnwidth]{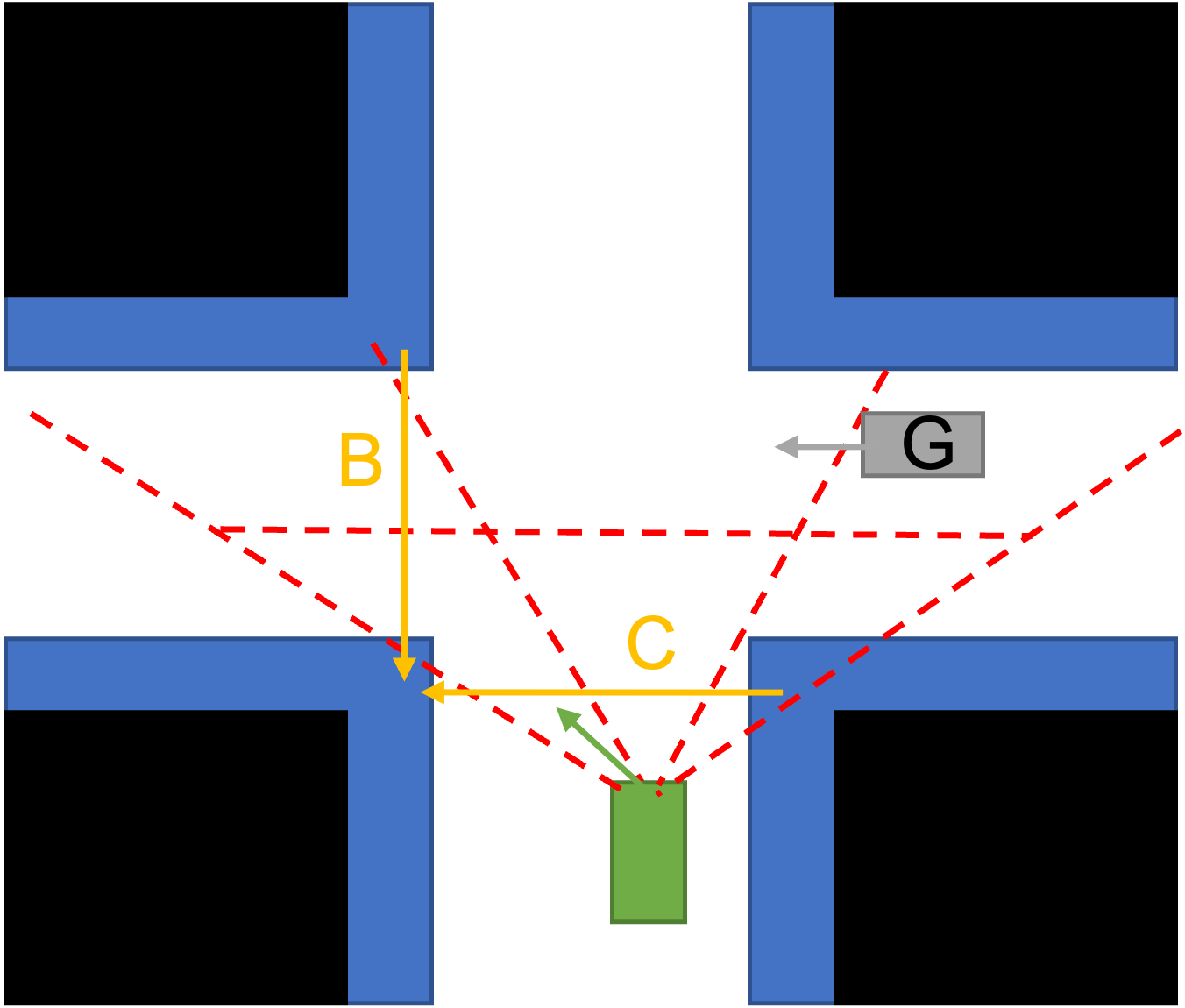}
    \caption{Left event intersection}
    \label{fig:events_l}
\end{subfigure}
\hfill
\begin{subfigure}{0.64\columnwidth}
    \centering
    \includegraphics[width=\columnwidth]{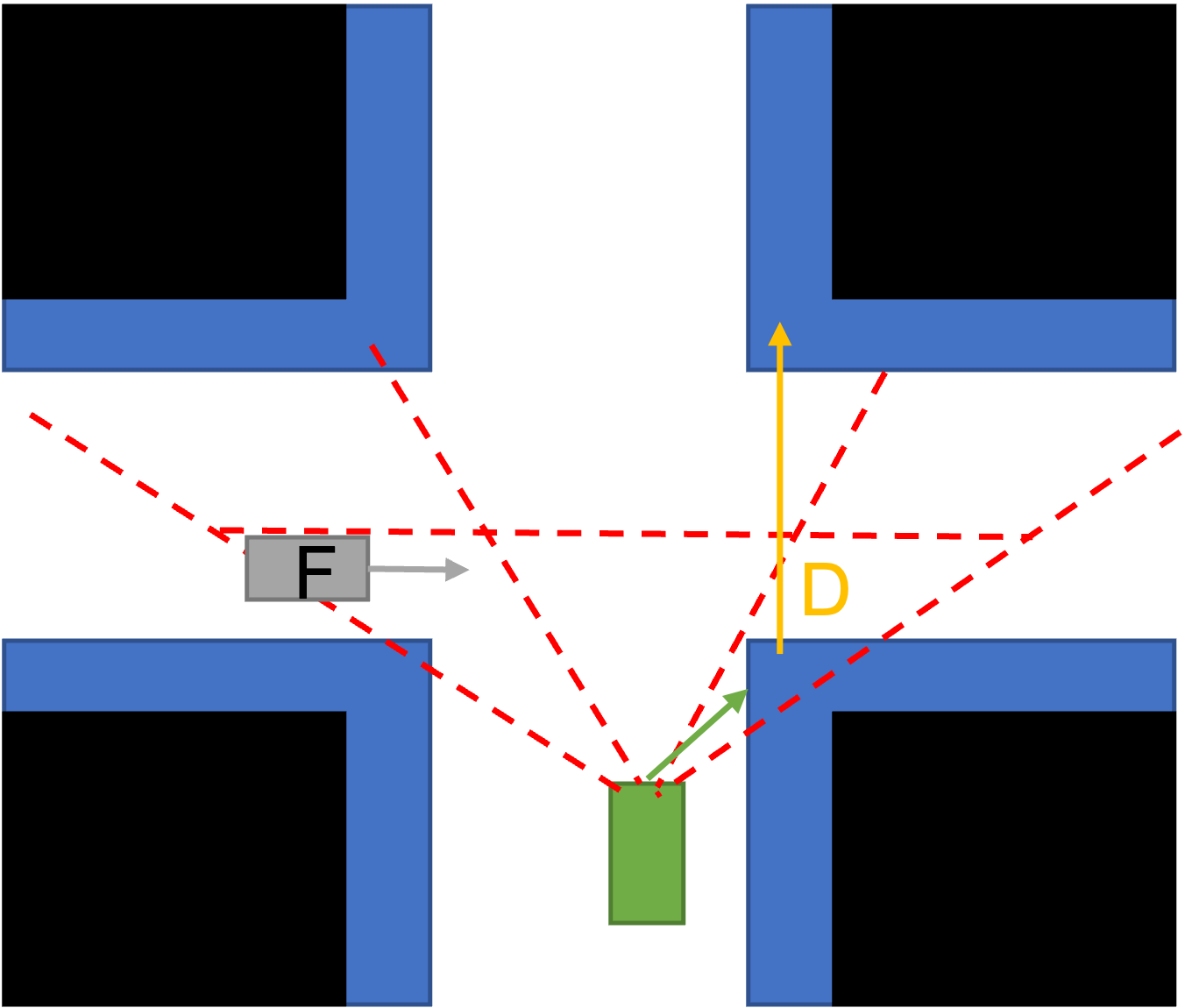}
    \caption{Right event intersection}
    \label{fig:events_r}
\end{subfigure}

\caption{We display the locations and heading directions of target objects in three types of event intersections. For clarity, distractor objects are not shown here. The green rectangle is the ego vehicle. Gray rectangles are other vehicles' locations and gray arrows show their moving directions. Yellows arrows show pedestrians' movements across the intersection.}
\label{fig:events}
\vspace{-10pt}
\end{figure*}

We inserted a SAGAT pause while the ego-car is waiting for other traffic at an intersection. The participant is asked to answer the positions of objects, including cars and pedestrians (Level 1 SA). We are particularly interested to study the SA of some of the objects that can potentially collide with the ego vehicle (the future trajectory of the object will intersect with the ego-car's future trajectory). We carefully design the event timing so that only these objects are located in certain regions at the SAGAT pauses. We refer to them as target objects and other objects as distractor objects. During the SAGAT pause, the simulation is frozen and other situational objects (i.e. pedestrians, vehicles and traffic lights) are hidden in the simulator. Meanwhile, several regions would be displayed on the blank scene, as shown in \Cref{fig:sagat}. The driver is asked to speak about all the regions where he/she believes there were pedestrians and/or vehicles. 
In particular, we have designed three types of events that correspond to three heading directions (i.e. forward, turning left and turning right) of the ego vehicle. 

Two driving routes with opposite directions and different traffic densities were designed in this experiment (See \Cref{fig:drives}). We change the traffic density of routes, by adding or removing distractor objects from event intersections. The average number of total objects, including both distractors and target objects, is 10 for {\it dense traffic} (DT) route and 5 for {\it light traffic} (LT) route drives in event intersections. Based on each route, two drives with different event designs were developed (LT1 \& LT2 and DT1 \& DT2).
\Cref{fig:events} illustrates the design of each type of event and target objects in the event:

\begin{itemize}[leftmargin=*]
    \item Forward intersections (a1 and a2 in \Cref{fig:drives}). For intersections where the ego vehicle is heading straight, the target objects are pedestrian A, vehicles F and G (\Cref{fig:events_f}). The SAGAT pause occurs while pedestrian A is crossing the intersection on the top (at area 1) and vehicles F and G are going through the intersection in the middle (at area 1 and area 2 respectively).    
    
    \item Left intersections (b1 and b2 in \Cref{fig:drives}). For intersections where the ego vehicle is heading left, the target objects are pedestrians B, C and vehicle G (\Cref{fig:events_l}). The SAGAT pause occurs while pedestrian B is crossing the intersection on the top left (at area 4), C is on the bottom center (at area 2) and vehicle G is waiting on the top right (at area 4).  
    
    \item Right intersections (c1 and c2 in \Cref{fig:drives}). For intersections where the ego vehicle is heading right, the target objects include the pedestrian D and vehicle F (\Cref{fig:events_r}). The SAGAT pause occurs while vehicle F is waiting on the bottom left (at area 3) and pedestrian D is crossing the intersection on the bottom right (at area 3).
\end{itemize}

Each participant was randomly assigned to one of the four experimental groups and experienced one of the four combinations of two drives (routes) in sequential order: group (i) LT1 and then DT2; group (ii) DT2 and then LT1; group (iii) LT2 and then DT1; and group (iv) DT1 and then LT2. \Cref{fig:drives} illustrates the event/non-event intersections for all four drives. For each drive, there are six event intersections, including two left intersections, two right intersections and two forward intersections. In some event intersections (blue dots in \Cref{fig:drives}), the target objects are highlighted. In order to reduce the learning effect between the two intersections of the same type within a drive, we also design a non-event intersection (green dots) between them: in these non-event intersections, the driver also needs to answer a dummy SAGAT question, such as the heading direction of the vehicle and the color of the traffic light. The average duration of a drive is 15 minutes.

Our goal is to understand how highlighting would change the SA and attention for different objects. Thus we measure SA from the same driver at two different timings in one type of intersection: 1) before treatment (i.e. highlighting or not highlighting the target object) and 2) 1 second after the treatment. To reduce the order effect, we implemented two separate intersections with the same type of events, but with different timing of the SAGAT pauses. For example, in drive LT1, the SAGAT pause in one of the forward event intersections (a2) is delayed by 1 second, while the SAGAT pause in the other forward event intersection (a1) is not delayed. The purpose of this delay is to study how highlighting or not highlighting the target object within this delayed period would change drivers' SA. Since the delayed and undelayed intersections have exactly the same event, we can compare the driver's SA responses to better understand the effect of highlighting.

\begin{figure*}
    \centering
    \begin{subfigure}{0.75\textwidth}
        \centering
        \includegraphics[width=\textwidth]{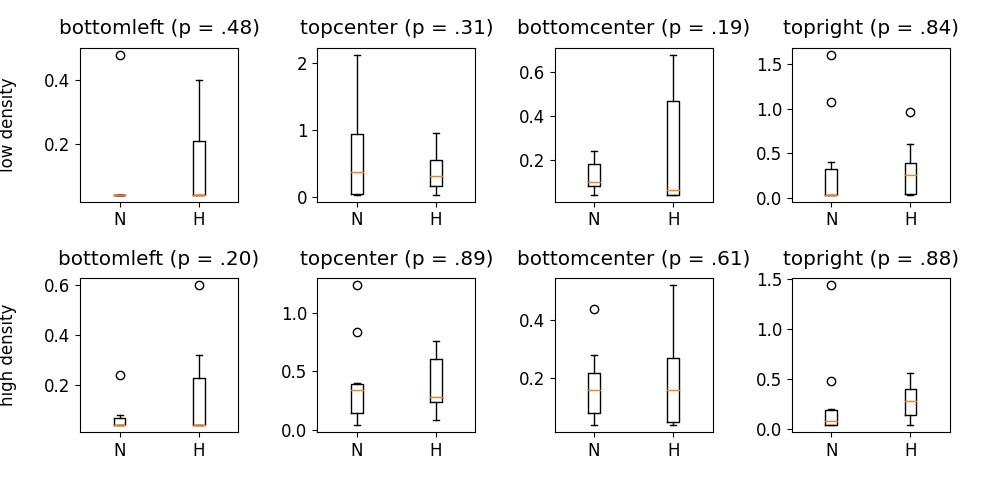}
        \caption{Fixation time on cars}
        \label{fig:fixation_car}
    \end{subfigure}
    \quad
    \begin{subfigure}{0.75\textwidth}
        \centering
        \includegraphics[width=\textwidth]{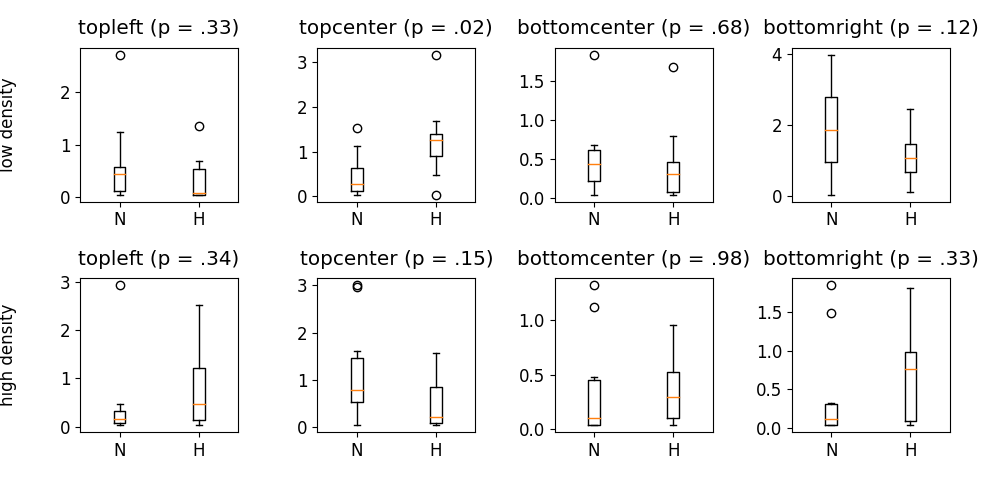}
        \caption{Fixation time on pedestrians}
        \label{fig:fixation_ped}
    \end{subfigure}
    \caption{Drivers' fixation time (in second) on each car and pedestrian given traffic density. ``N'' represents non-highlighting results and ``H'' represents highlighting results. We report the p-value between highlighting conditions for each object.}
    \label{fig:fixation}
\end{figure*}

\subsection{Dependent Variables}
\label{sec:dependent}
\noindent \textbf{Attention allocation.} Attention allocation is strongly associated with situational awareness. To form situational awareness, one needs to perceive and process the environment \cite{endsley1988design}. However, the limited capacity of human attentional resources in combination with the excessive attentional demands in a dynamic driving environment can result in a loss of situational awareness. To study attention allocation, one well-established measurement is to track human fixation behavior. We collect drivers' eye-tracking data with Tobii Pro Glasses 3. We also annotate the target objects in each event intersection using Vatic \cite{vondrick2013efficiently}. Based on the finding that humans can recognize information in the fovea (2.5 deg \cite{nelson1980functional, quinn2019clinical}) within 120 ms \cite{rayner2007eye}, we define that the driver has fixated on an object if the gaze has stayed within 2.5 degrees from the center of the object for more than 120 milliseconds.

\begin{figure*}
    \centering
    \begin{subfigure}{0.75\textwidth}
        \centering
        \includegraphics[width=\textwidth]{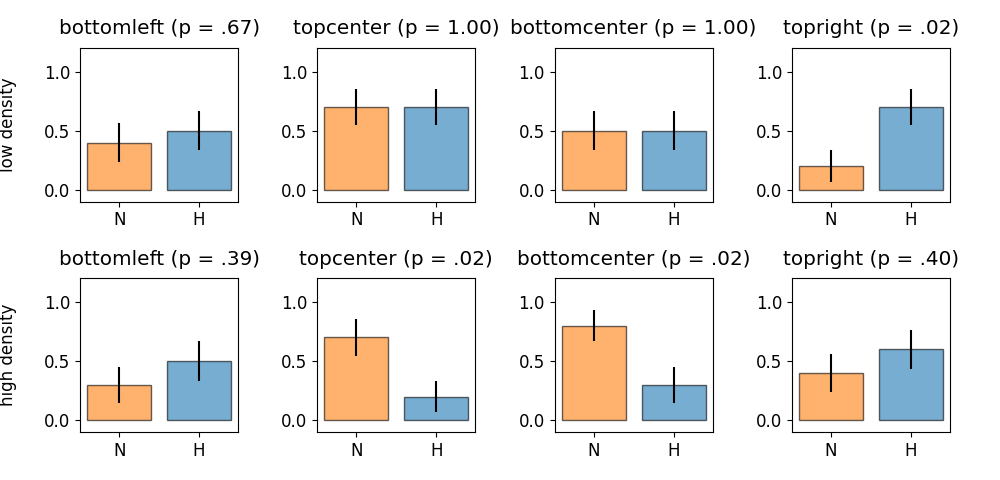}
        \caption{Response accuracy on cars}
        \label{fig:sa_car}
    \end{subfigure}
    \quad
    \begin{subfigure}{0.75\textwidth}
        \centering
        \includegraphics[width=\textwidth]{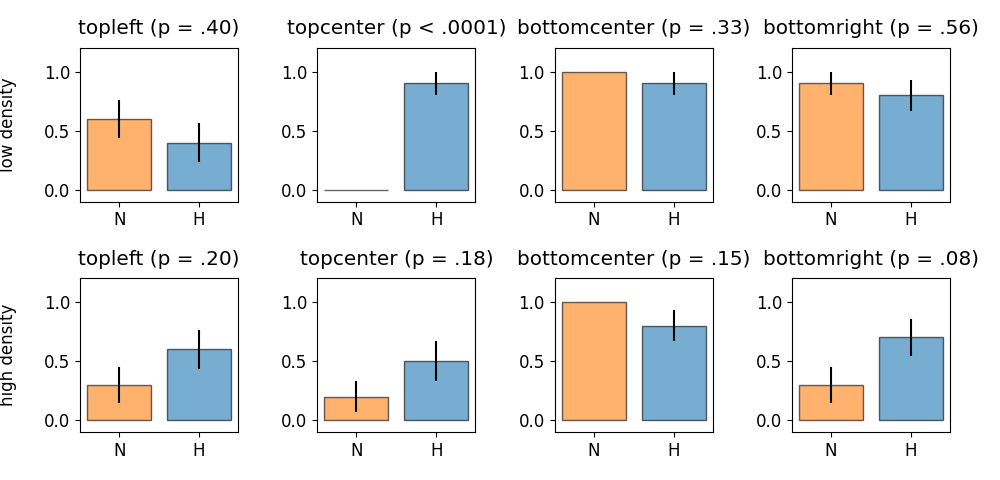}
        \caption{Response accuracy on pedestrians}
        \label{fig:sa_ped}
    \end{subfigure}
    \caption{Drivers' SAGAT question response accuracy in delayed intersections. ``N'' represents non-highlighting results and ``H'' represents highlighting results. We report the p-value between highlighting conditions for each object.}
    \label{fig:sa_accuracy}
    \vspace{-10pt}
\end{figure*}

\begin{figure}[t]
    \centering 
    \begin{subfigure}{.49\columnwidth}
      \includegraphics[width=\columnwidth]{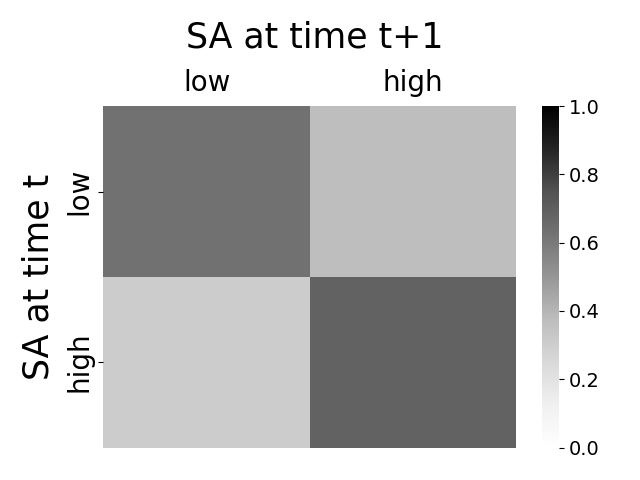}
      \caption{low density w/o highlighting}
    \end{subfigure} 
    \begin{subfigure}{.49\columnwidth}
      \includegraphics[width=\columnwidth]{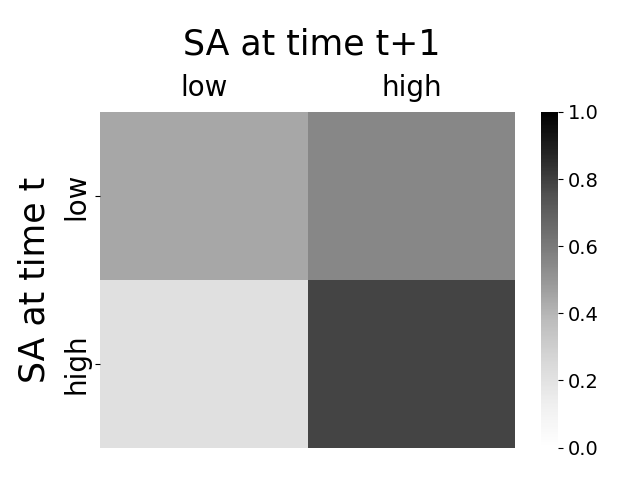}
      \caption{low density w/ highlighting}
    \end{subfigure} 
    
    \medskip
    \begin{subfigure}{.49\columnwidth}
      \includegraphics[width=\columnwidth]{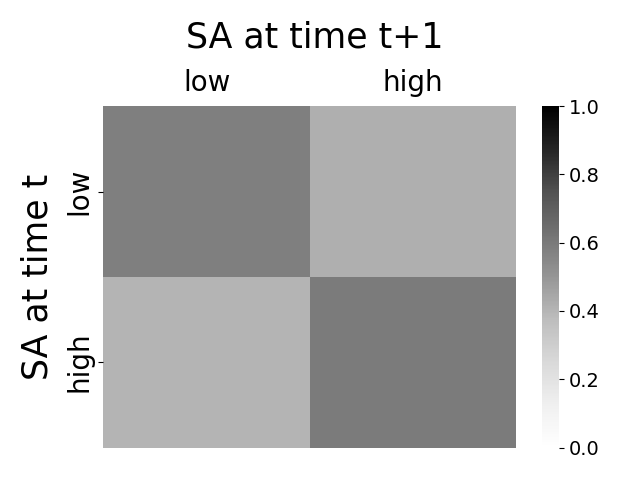}
      \caption{high density w/o highlighting}
    \end{subfigure}
    \begin{subfigure}{.49\columnwidth}
      \includegraphics[width=\columnwidth]{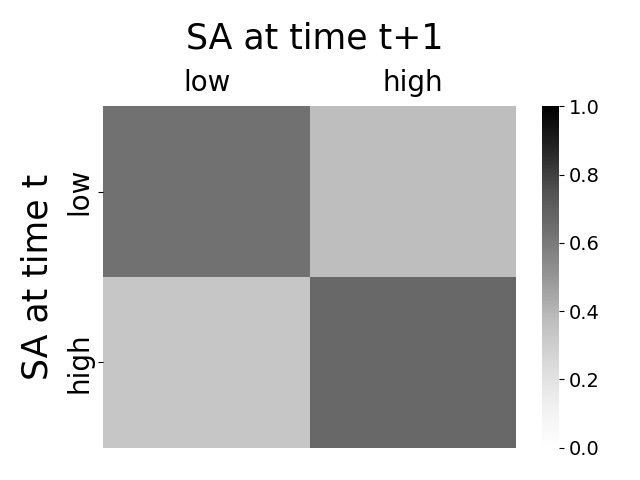}
      \caption{high density w highlighting}
    \end{subfigure}
\caption{SA transition conditioned on traffic density and highlighting across all objects. "SA at time t" represents drivers' SA response before the treatment, while "SA at time t+1" is for SA response after the treatment. The shade of each region represents the proportion of the samples falling into each category. Darker color represents a higher proportion.}
\label{fig:SA_transition_all}
\end{figure}

\begin{figure}[t]
    \centering 
    \begin{subfigure}{.49\columnwidth}
      \includegraphics[width=\columnwidth]{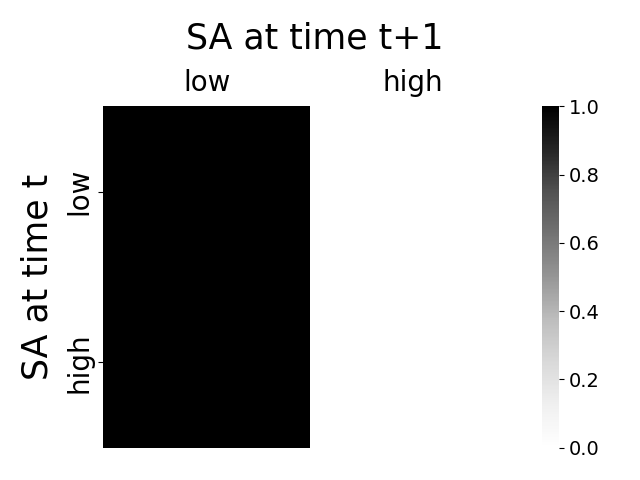}
      \caption{low density w/o highlighting}
    \end{subfigure} 
    \begin{subfigure}{.49\columnwidth}
      \includegraphics[width=\columnwidth]{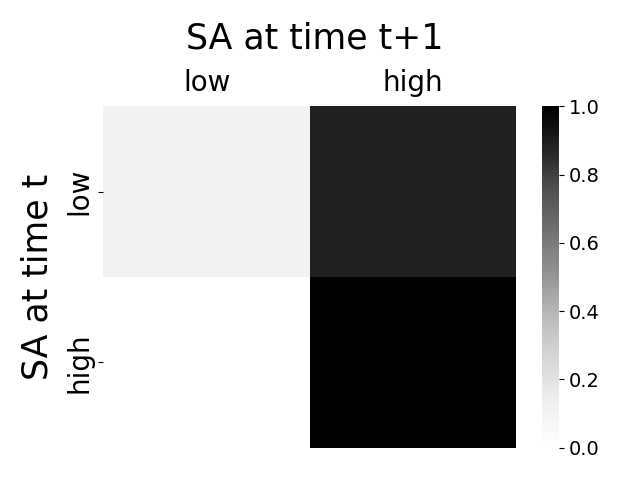}
      \caption{low density w/ highlighting}
    \end{subfigure} 
    
    \medskip
    \begin{subfigure}{.49\columnwidth}
      \includegraphics[width=\columnwidth]{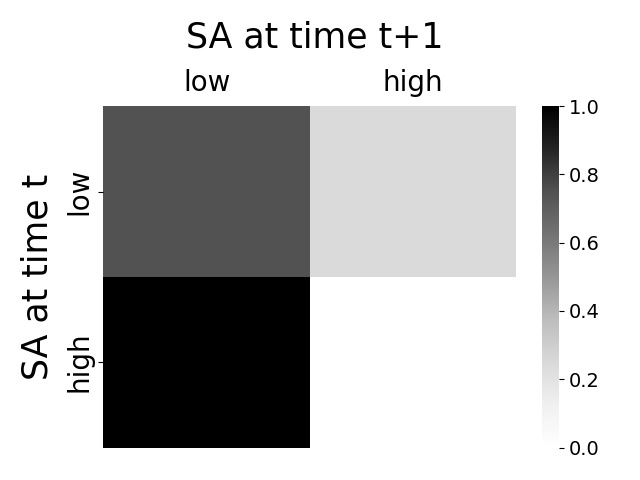}
      \caption{high density w/0 highlighting}
    \end{subfigure}
    \begin{subfigure}{.49\columnwidth}
      \includegraphics[width=\columnwidth]{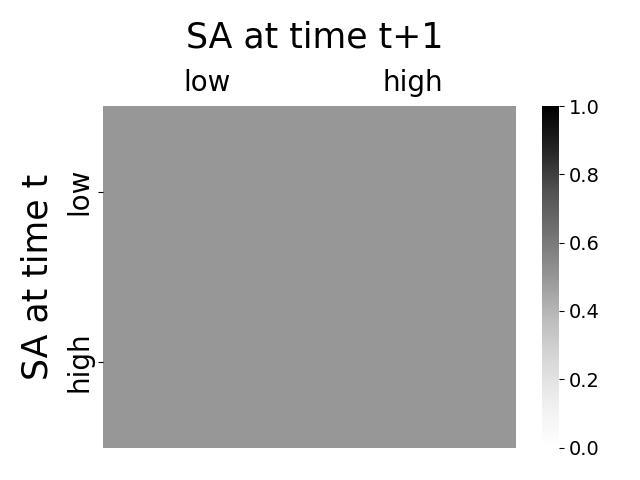}
      \caption{high density w/ highlighting}
    \end{subfigure}
    
\caption{SA transition for the top center pedestrian (pedestrian A in \Cref{fig:events_f}). The shade of each region represents the proportion of the samples falling into each category. Darker color represents a higher proportion.}
\label{fig:SA_transition_topcenter_ped}
\end{figure}

\begin{figure}[t]
    \centering 
    \begin{subfigure}{.49\columnwidth}
      \includegraphics[width=\columnwidth]{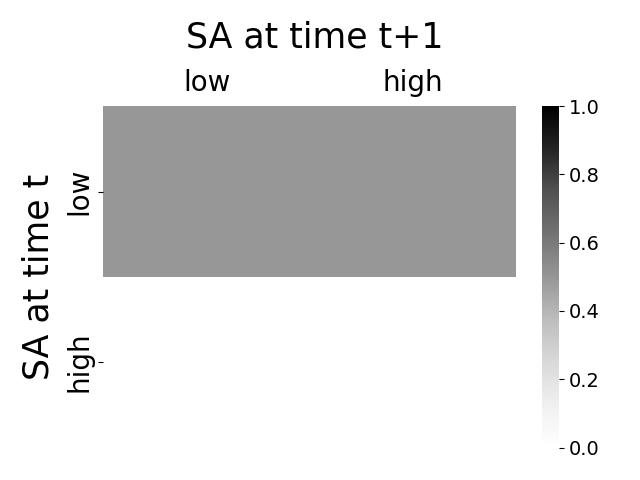}
      \caption{low density w/o highlighting}
    \end{subfigure} 
    \begin{subfigure}{.49\columnwidth}
      \includegraphics[width=\columnwidth]{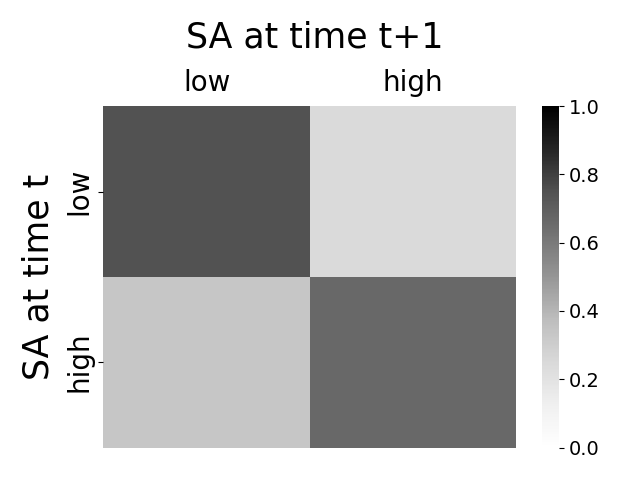}
      \caption{low density w/ highlighting}
    \end{subfigure} 
    
    \medskip
    \begin{subfigure}{.49\columnwidth}
      \includegraphics[width=\columnwidth]{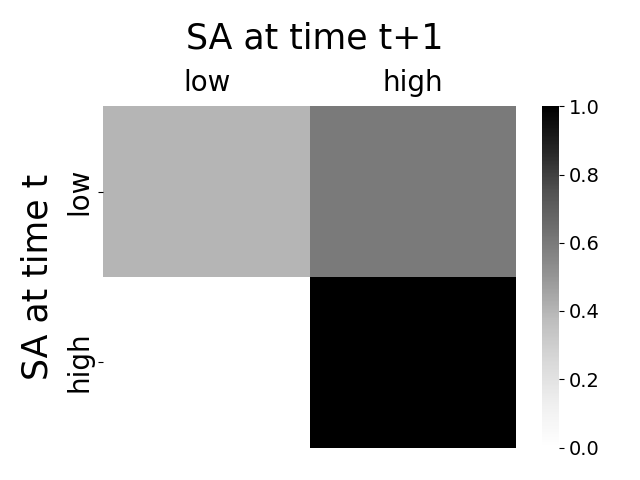}
      \caption{high density w/o highlighting}
    \end{subfigure}
    \begin{subfigure}{.49\columnwidth}
      \includegraphics[width=\columnwidth]{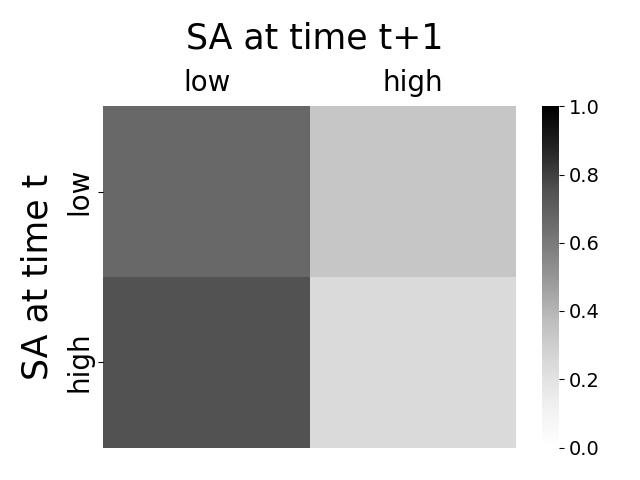}
      \caption{high density w/ highlighting}
    \end{subfigure}
    
\caption{SA transition for the bottom center car (car F in \Cref{fig:events_f}). The shade of each region represents the proportion of the samples falling into each category. Darker color represents a higher proportion.}
\label{fig:SA_transition_bottomcenter_car}
\end{figure}

\noindent \textbf{Situational awareness.} 
To measure drivers' situational awareness, we adopt the SAGAT technique and ask drivers questions about the location of pedestrians and vehicles during the pauses in the event intersections (\Cref{fig:sagat}). We focus on the Level 1 SA (perception) since it is the most fundamental one. Participants were asked to select all the regions that they think had a pedestrian or a vehicle the moment before the SAGAT pause. Two images with region highlightings, each corresponding to pedestrians and vehicles (e.g., \Cref{fig:sagat} is for pedestrians), are presented in sequence with a corresponding SAGAT question. They were also asked to provide a confidence level (from 0 to 100) for each region they specified, which is further discretized to low confidence (0-50) and high confidence (51-100). Regions that are not selected by the participant are treated as low confidence. The order of SAGAT questions (i.e. pedestrian or vehicle) is randomized to reduce the order effect. During the analysis, we study the SA response on the discretized regions that are occupied by the target objects.

\subsection{Procedure}
\label{sec:procedure}
Participants first completed a pre-study survey to provide demographics and driving experience. They also filled in a questionnaire designed to evaluate their trust in automation \cite{jian2000foundations}. The moderator then gave each participant a brief introduction to the study and set up the Tobii glasses and the driving simulator. The study began with a practice drive where the participant was asked a sample question during a SAGAT pause. During the practice drive, the participant was asked to take over control of the vehicle using the brake pedal whenever they felt uncomfortable with the AD system. The participant was also given a chance to practice answering the SAGAT question at an intersection as well as indicating their intention to take over. After the practice drive, the participant was randomly assigned to an experimental group, and went through two standard drives of different traffic densities. During a drive, each participant experienced six event intersections and two non-event intersections (\Cref{fig:drives}). In each event intersection, the participant was asked about the locations of vehicles and pedestrians during the SAGAT pause (\Cref{fig:sagat}). 

\section{Results}
In this section, we present the results of our study, analyzing how highlighting objects changes drivers' attention allocation and situational awareness during their interaction with an AD system based on the data collected from the driving simulator experiment. 

\subsection{Attention Allocation}
For attention allocation, we analyze the driver's fixation time on target objects. Since our goal is to study how highlighting would change the driver's attention, we focus on the driver's fixation during the delayed period (\Cref{sec:scenarios}). We show the results for specific cars and pedestrians in \Cref{fig:fixation} at different traffic densities. Running a pairwise t-test, we found a significant effect ($p=.02$) for highlighting the top center pedestrian, i.e. the pedestrian A, when the traffic density is low. We don't find the same trend for top center pedestrians at high traffic density.

\subsection{Situational Awareness}
\noindent \textbf{SA response accuracy.} 
We analyze drivers' responses to the SAGAT questions at delayed intersections, when different highlighting conditions have been applied to the target objects (\Cref{fig:sa_accuracy}). Across all objects, driver's SA on highlighted objects ($M=0.60, SD=0.49$) are higher than the unhighlighted ones ($M=0.52, SD=0.50$), but the difference is not statistically significant ($p=.14$). For cars (\Cref{fig:sa_car}), we observed a significant difference between highlighting conditions for the top right car (car G) in a low traffic density environment($p=.02$) and for the top center car and bottom center car (cars F and G) in high traffic density environment($p=.02$). For pedestrians (\Cref{fig:sa_ped}), we only found a significant change in SA for top center pedestrians during light traffic routes ($p<.0001$). The results indicate that highlighting can improve the SA for the top right car and the top center pedestrian at low traffic density, while decreasing the SA for the bottom center car and top center car at high traffic density. The Pearson correlation coefficient between fixation time and SA response accuracy is $r=.12$ ($p=.03$), indicating a weak correlation between attention and SA.

\noindent \textbf{SA transition.}
We first analyze the transition of drivers' SA from undelayed pauses to delayed pauses across all objects, when we apply different highlighting conditions to the target objects during time $t$ and time $t+1$ (\Cref{fig:SA_transition_all}). Given low traffic density, for drivers with an initial low SA on target objects, highlighting leads to SA improvement (from low to high) for 55.3\% of the drivers, compared to 36.8\% in the non-highlighting conditions. For drivers with an initial high SA on target objects, we found that those in the highlighting conditions are more likely to maintain their high SA (78.6\%) compared to drivers in the non-highlighting conditions (69.0\%) for low density. Similarly, when the traffic density is high, highlighting also helps more drivers maintain high SA (66.7\%) compared to the no highlighting (59.5\%). Running a two-sample proportion test, however, we didn't find any significant effect of highlighting on SA transition for either traffic density across all objects. 

Looking at the SA transition for specific objects, we found from a proportion test that for the top center pedestrian, highlighting can significantly increase the proportion of drivers that improve low SA ($p=.0007$) and maintain high SA ($p=.03$) compared to the control condition when the traffic density is low (\Cref{fig:SA_transition_topcenter_ped}). On the contrary, for the bottom center car, we found that highlighting actually decreases the proportion of drivers that maintain high SA ($p=.02$) when the traffic density is high (\Cref{fig:SA_transition_bottomcenter_car}). We didn't find any significant difference in SA transition between highlighting conditions for other objects. 

\section{Discussion}
The results indicate that the effect of highlighting varies a lot depending on the situation. Highlighting can significantly improve SA on certain objects at low traffic density. However, it can also decrease drivers' SA of some objects at high traffic density. These findings can provide guidance in selecting which object to highlight for the UI to improve the driver's SA while driving and monitoring SAE L2 or L3 AVs.

\subsection{Attention Allocation, Workload and SA}
Driving is a visual and motor control process. Thus, drivers' attention allocation and workload play important roles in establishing their SA. Previous works have proposed quantitative methods to model the interplay between attention allocation, workload and SA \cite{wickens2008attention, shuang2014quantitative, liu2014modeling}than on unhighlighted ones. Specifically, the attention allocation process can be largely influenced by the salience of an object and workload \cite{wickens2002situation}. A high workload can result in attention tunneling and negatively impact SA. In our study results, highlighting the cars in the center of the driver's field of view significantly decreases SA when the traffic density is high, while the difference is not significant when the traffic density is lower. This can probably be explained by (i) the driver's high workload given the dense traffic ii) the highlighting AR cues induce additional workload (iii) the fact that cars in the center are already very salient even without highlighting. These reasons can also explain why significant SA improvement was found for the top center pedestrian (which is not visually salient and easy to be ignored by the driver) at low traffic density and why improvement is not significant at higher traffic density (due to the driver being overwhelmed by the dense traffic). We believe these results shed light on designing object-specific AR cues on human-machine interfaces. 

\subsection{Comparison with Previous Studies}
Previous works focus on evaluating the driver's average SA across all traffic objects in different experimental groups. By controlling a specific object's spatial characteristic in the driving simulator, we are able to further study the transition of the user's SA on the object before and after the highlighting. Results from previous works \cite{phan2016enhancing, lindemann2018catch, colley2021effects} showed that using an AR interface could improve drivers' average SA across all objects. 
Thanks to the unique study design covering objects properties and traffic conditions in common intersections as well as the proposed SAGAT protocol with temporal variations, we are able to see significant positive effects of AR cues on some objects and negative effects on some other objects. These results extend knowledge of the community on the effects of AR cues beyond specifically-designed scenarios and hand-picked objects, showing how different objects can benefit from the AR cues in more general driving scenarios.

\subsection{Limitations and Future Work}
Our UI is implemented in a driving simulator, which enables us to control the timing of events accurately. Driving scenarios in the real world are more complex and have more variety than our examined scenarios. In reality, the AR cues can be implemented by detecting vehicles and pedestrians from sensors and highlighting them using bounding boxes on the AR-HUD. In addition, every participant experienced two similar event intersections - one before and one after the highlighting. We ask dummy SAGAT questions in non-event intersections between the two events intersections to reduce the learning effect, but the effect may not be canceled off completely. Additionally, we measure SA using SAGAT, which is known to be highly reliable \cite{endsley1994individual}. The drawback is that SAGAT requires the participant to memorize the objects and thus can also increase the workload \cite{fujino2020comparison}. Non-intrusive SA measures can be considered in a future study to ensure an accurate measure of drivers' workload when interacting with an AD system. Finally, in the future work, we plan to consider other object features (e.g. object colors and speed) and the differences in the intersections' background environment, which are also likely to affect SA.

\section{Conclusion}
This work aims to investigate the effects of highlighting objects with an AR interface on drivers' perception-level SA for SAE L2 or L3 AVs under different circumstances, including object types, locations and traffic densities in urban environments. We conducted a user study in a driving simulator ($N=20$). The results show that highlighting has a positive impact on SA when the traffic density is low and the highlighted object has originally low visual saliency, and sometimes causes a reduction in SA when the object is already very salient even without highlighting during dense traffic. This work extends the knowledge on methods to improve driver's situational awareness for autonomous vehicles, and enables the development of a smart driver-assistance interface that can selectively highlight objects to improve SA for drivers monitoring partially autonomous vehicles.

\bibliographystyle{IEEEtran}
\bibliography{IEEEexample}

\begin{thebibliography}{10}
\providecommand{\url}[1]{#1}
\csname url@rmstyle\endcsname
\providecommand{\newblock}{\relax}
\providecommand{\bibinfo}[2]{#2}
\providecommand\BIBentrySTDinterwordspacing{\spaceskip=0pt\relax}
\providecommand\BIBentryALTinterwordstretchfactor{4}
\providecommand\BIBentryALTinterwordspacing{\spaceskip=\fontdimen2\font plus
\BIBentryALTinterwordstretchfactor\fontdimen3\font minus
  \fontdimen4\font\relax}
\providecommand\BIBforeignlanguage[2]{{%
\expandafter\ifx\csname l@#1\endcsname\relax
\typeout{** WARNING: IEEEtran.bst: No hyphenation pattern has been}%
\typeout{** loaded for the language `#1'. Using the pattern for}%
\typeout{** the default language instead.}%
\else
\language=\csname l@#1\endcsname
\fi
#2}}

\bibitem{sae2018taxonomy}
S.~international, ``Taxonomy and definitions for terms related to driving
  automation systems for on-road motor vehicles,'' \emph{SAE}, 2018.

\bibitem{endsley1988design}
M.~R. Endsley, ``Design and evaluation for situation awareness enhancement,''
  in \emph{Proceedings of the Human Factors Society annual meeting}, vol.~32,
  no.~2.\hskip 1em plus 0.5em minus 0.4em\relax Sage Publications Sage CA: Los
  Angeles, CA, 1988, pp. 97--101.

\bibitem{helldin2014transparency}
T.~Helldin, ``Transparency for future semi-automated systems: Effects of
  transparency on operator performance, workload and trust,'' Ph.D.
  dissertation, {\"O}rebro Universitet, 2014.

\bibitem{ananny2018seeing}
M.~Ananny and K.~Crawford, ``Seeing without knowing: Limitations of the
  transparency ideal and its application to algorithmic accountability,''
  \emph{new media \& society}, vol.~20, no.~3, pp. 973--989, 2018.

\bibitem{lindemann2018catch}
P.~Lindemann, T.-Y. Lee, and G.~Rigoll, ``Catch my drift: Elevating situation
  awareness for highly automated driving with an explanatory windshield display
  user interface,'' \emph{Multimodal Technologies and Interaction}, vol.~2,
  no.~4, p.~71, 2018.

\bibitem{colley2021effects}
M.~Colley, B.~Eder, J.~O. Rixen, and E.~Rukzio, ``Effects of semantic
  segmentation visualization on trust, situation awareness, and cognitive load
  in highly automated vehicles,'' in \emph{Proceedings of the 2021 CHI
  Conference on Human Factors in Computing Systems}, 2021, pp. 1--11.

\bibitem{durso1998situation}
F.~T. Durso, C.~A. Hackworth, T.~R. Truitt, J.~Crutchfield, D.~Nikolic, and
  C.~A. Manning, ``Situation awareness as a predictor of performance for en
  route air traffic controllers,'' \emph{Air Traffic Control Quarterly},
  vol.~6, no.~1, pp. 1--20, 1998.

\bibitem{sirkin2017toward}
D.~Sirkin, N.~Martelaro, M.~Johns, and W.~Ju, ``Toward measurement of situation
  awareness in autonomous vehicles,'' in \emph{Proceedings of the 2017 CHI
  Conference on Human Factors in Computing Systems}, 2017, pp. 405--415.

\bibitem{taylor1990situational}
R.~Taylor, ``Situational awareness rating technique (sart): The development of
  a tool for aircrew systems design. situational awareness in aerospace
  operations (agard-cp-478),'' \emph{Neuilly Sur Seine, France: NATO-AGARD},
  1990.

\bibitem{waag1994tools}
W.~L. Waag and M.~R. Houck, ``Tools for assessing situational awareness in an
  operational fighter environment.'' \emph{Aviation, space, and environmental
  medicine}, 1994.

\bibitem{endsley1988situation}
M.~R. Endsley, ``Situation awareness global assessment technique (sagat),'' in
  \emph{Proceedings of the IEEE 1988 national aerospace and electronics
  conference}.\hskip 1em plus 0.5em minus 0.4em\relax IEEE, 1988, pp. 789--795.

\bibitem{endsley1994individual}
M.~R. Endsley and C.~A. Bolstad, ``Individual differences in pilot situation
  awareness,'' \emph{The International Journal of Aviation Psychology}, vol.~4,
  no.~3, pp. 241--264, 1994.

\bibitem{endsley1990predictive}
M.~R. Endsley, ``Predictive utility of an objective measure of situation
  awareness,'' in \emph{Proceedings of the Human Factors Society annual
  meeting}, vol.~34, no.~1.\hskip 1em plus 0.5em minus 0.4em\relax SAGE
  Publications Sage CA: Los Angeles, CA, 1990, pp. 41--45.

\bibitem{barnard2010spotting}
Y.~Barnard and F.~Lai, ``Spotting sheep in yorkshire: Using eye-tracking for
  studying situation awareness in a driving simulator,'' in \emph{Human
  Factors: A System View of Human, Technology and Organisation. Annual
  Conference of the Europe Chapter of the Human Factors and Ergonomics Society
  2009}, 2010.

\bibitem{yang2018hmi}
Y.~Yang, B.~Karakaya, G.~C. Dominioni, K.~Kawabe, and K.~Bengler, ``An hmi
  concept to improve driver's visual behavior and situation awareness in
  automated vehicle,'' in \emph{2018 21st International Conference on
  Intelligent Transportation Systems (ITSC)}.\hskip 1em plus 0.5em minus
  0.4em\relax IEEE, 2018, pp. 650--655.

\bibitem{Zhu2021ImprovingDS}
H.~Zhu, T.~Misu, S.~Martin, X.~Wu, and K.~Akash, ``Improving driver situation
  awareness prediction using human visual sensory and memory mechanism,''
  \emph{arXiv preprint arXiv:2111.00087}, 2021.

\bibitem{salmon2006situation}
P.~Salmon, N.~Stanton, G.~Walker, and D.~Green, ``Situation awareness
  measurement: A review of applicability for c4i environments,'' \emph{Applied
  ergonomics}, vol.~37, no.~2, pp. 225--238, 2006.

\bibitem{bolstad2001situation}
C.~A. Bolstad, ``Situation awareness: does it change with age?'' in
  \emph{Proceedings of the human factors and ergonomics society annual
  meeting}, vol.~45, no.~4.\hskip 1em plus 0.5em minus 0.4em\relax SAGE
  Publications Sage CA: Los Angeles, CA, 2001, pp. 272--276.

\bibitem{wright2016experienced}
T.~J. Wright, S.~Samuel, A.~Borowsky, S.~Zilberstein, and D.~L. Fisher,
  ``Experienced drivers are quicker to achieve situation awareness than
  inexperienced drivers in situations of transfer of control within a level 3
  autonomous environment,'' in \emph{Proceedings of the Human Factors and
  Ergonomics Society Annual Meeting}, vol.~60, no.~1.\hskip 1em plus 0.5em
  minus 0.4em\relax Sage Publications Sage CA: Los Angeles, CA, 2016, pp.
  270--273.

\bibitem{johannsdottir2010role}
K.~R. Johannsdottir and C.~M. Herdman, ``The role of working memory in
  supporting drivers’ situation awareness for surrounding traffic,''
  \emph{Human factors}, vol.~52, no.~6, pp. 663--673, 2010.

\bibitem{heenan2014effects}
A.~Heenan, C.~M. Herdman, M.~S. Brown, and N.~Robert, ``Effects of conversation
  on situation awareness and working memory in simulated driving,'' \emph{Human
  factors}, vol.~56, no.~6, pp. 1077--1092, 2014.

\bibitem{merenda2018augmented}
C.~Merenda, H.~Kim, K.~Tanous, J.~L. Gabbard, B.~Feichtl, T.~Misu, and C.~Suga,
  ``Augmented reality interface design approaches for goal-directed and
  stimulus-driven driving tasks,'' \emph{IEEE transactions on visualization and
  computer graphics}, vol.~24, no.~11, pp. 2875--2885, 2018.

\bibitem{wang2020situation}
J.~Wang, W.~Wang, P.~Hansen, Y.~Li, and F.~You, ``The situation awareness and
  usability research of different hud hmi design in driving while using
  adaptive cruise control,'' in \emph{International Conference on
  Human-Computer Interaction}.\hskip 1em plus 0.5em minus 0.4em\relax Springer,
  2020, pp. 236--248.

\bibitem{tong2019augmented}
Y.~Tong and B.~Jia, ``An augmented-reality-based warning interface for
  pedestrians: User interface design and evaluation,'' in \emph{Proceedings of
  the Human Factors and Ergonomics Society Annual Meeting}, vol.~63,
  no.~1.\hskip 1em plus 0.5em minus 0.4em\relax SAGE Publications Sage CA: Los
  Angeles, CA, 2019, pp. 1834--1838.

\bibitem{phan2016enhancing}
M.~T. Phan, I.~Thouvenin, and V.~Fr{\'e}mont, ``Enhancing the driver awareness
  of pedestrian using augmented reality cues,'' in \emph{2016 IEEE 19th
  International Conference on Intelligent Transportation Systems (ITSC)}.\hskip
  1em plus 0.5em minus 0.4em\relax IEEE, 2016, pp. 1298--1304.

\bibitem{shah2018airsim}
S.~Shah, D.~Dey, C.~Lovett, and A.~Kapoor, ``Airsim: High-fidelity visual and
  physical simulation for autonomous vehicles,'' in \emph{Field and service
  robotics}.\hskip 1em plus 0.5em minus 0.4em\relax Springer, 2018, pp.
  621--635.

\bibitem{nelson1980functional}
W.~W. Nelson and G.~R. Loftus, ``The functional visual field during picture
  viewing.'' \emph{Journal of Experimental Psychology: Human Learning and
  Memory}, vol.~6, no.~4, p. 391, 1980.

\bibitem{quinn2019clinical}
N.~Quinn, L.~Csincsik, E.~Flynn, C.~A. Curcio, S.~Kiss, S.~R. Sadda, R.~Hogg,
  T.~Peto, and I.~Lengyel, ``The clinical relevance of visualising the
  peripheral retina,'' \emph{Progress in retinal and eye research}, vol.~68,
  pp. 83--109, 2019.

\bibitem{carrasco1995eccentricity}
M.~Carrasco, D.~L. Evert, I.~Chang, and S.~M. Katz, ``The eccentricity effect:
  Target eccentricity affects performance on conjunction searches,''
  \emph{Perception \& psychophysics}, vol.~57, no.~8, pp. 1241--1261, 1995.

\bibitem{vondrick2013efficiently}
C.~Vondrick, D.~Patterson, and D.~Ramanan, ``Efficiently scaling up
  crowdsourced video annotation,'' \emph{International journal of computer
  vision}, vol. 101, no.~1, pp. 184--204, 2013.

\bibitem{rayner2007eye}
K.~Rayner and M.~Castelhano, ``Eye movements,'' \emph{Scholarpedia}, vol.~2,
  no.~10, p. 3649, 2007.

\bibitem{jian2000foundations}
J.-Y. Jian, A.~M. Bisantz, and C.~G. Drury, ``Foundations for an empirically
  determined scale of trust in automated systems,'' \emph{International journal
  of cognitive ergonomics}, vol.~4, no.~1, pp. 53--71, 2000.

\bibitem{wickens2008attention}
C.~D. Wickens, J.~S. McCarley, A.~L. Alexander, L.~C. Thomas, M.~Ambinder, and
  S.~Zheng, ``Attention-situation awareness (a-sa) model of pilot error,''
  \emph{Human performance modeling in aviation}, pp. 213--239, 2008.

\bibitem{shuang2014quantitative}
L.~Shuang, W.~Xiaoru, and Z.~Damin, ``A quantitative situational awareness
  model of pilot,'' in \emph{Proceedings of the International Symposium on
  Human Factors and Ergonomics in Health Care}, vol.~3, no.~1.\hskip 1em plus
  0.5em minus 0.4em\relax SAGE Publications Sage CA: Los Angeles, CA, 2014, pp.
  117--122.

\bibitem{liu2014modeling}
S.~Liu, X.~Wanyan, and D.~Zhuang, ``Modeling the situation awareness by the
  analysis of cognitive process,'' \emph{Bio-medical materials and
  engineering}, vol.~24, no.~6, pp. 2311--2318, 2014.

\bibitem{wickens2002situation}
C.~D. Wickens, ``Situation awareness and workload in aviation,'' \emph{Current
  directions in psychological science}, vol.~11, no.~4, pp. 128--133, 2002.

\bibitem{fujino2020comparison}
M.~Fujino, J.~Lee, T.~Hirano, Y.~Saito, and M.~Itoh, ``Comparison of sagat and
  spam for seeking effective way to evaluate situation awareness and workload
  during air traffic control task,'' in \emph{Proceedings of the Human Factors
  and Ergonomics Society Annual Meeting}, vol.~64, no.~1.\hskip 1em plus 0.5em
  minus 0.4em\relax SAGE Publications Sage CA: Los Angeles, CA, 2020, pp.
  1836--1840.

\end{thebibliography}

\end{document}